\documentclass[prl,twocolumn,lengthcheck,superscriptaddress,showpacs,letterpaper,nofootinbib]{revtex4}

\usepackage{color}
\usepackage{graphicx}  
\usepackage{float}
\usepackage{amsmath}

\begin{document}

\title{Prospects for Detection of Gravitational
Waves from Intermediate-Mass-Ratio Inspirals}


\author{Duncan A. Brown}
\affiliation{LIGO Laboratory, California Institute of Technology, Pasadena, California 91125}
\affiliation{Theoretical Astrophysics, California Institute of Technology, Pasadena, California 91125}
\affiliation{Department of Physics, Syracuse University, Syracuse, New York 13244}
\author{Jeandrew Brink}
\affiliation{Theoretical Astrophysics, California Institute of Technology, Pasadena, California 91125}
\author{Hua Fang}
\affiliation{Theoretical Astrophysics, California Institute of Technology, Pasadena, California 91125}
\author{Jonathan R. Gair}
\affiliation{Institute of Astronomy, Madingley Road, Cambridge, CB3 0HA, UK}
\author{Chao Li}
\affiliation{Theoretical Astrophysics, California Institute of Technology, Pasadena, California 91125}
\author{Geoffrey Lovelace}
\affiliation{Theoretical Astrophysics, California Institute of Technology, Pasadena, California 91125}
\author{Ilya Mandel}
\affiliation{Theoretical Astrophysics, California Institute of Technology, Pasadena, California 91125}
\author{Kip S. Thorne}
\affiliation{Theoretical Astrophysics, California Institute of Technology, Pasadena, California 91125}
\begin{abstract} We explore prospects for detecting gravitational waves from
stellar-mass compact objects spiraling into intermediate mass black holes
(BHs) ($M\sim 50 M_\odot$ to $350 M_\odot$) with ground-based observatories.
We estimate an event rate for such \emph{intermediate-mass-ratio inspirals} of
$\lesssim 10$--$30\ \mathrm{yr}^{-1}$ in Advanced LIGO.  We show that if the
central body is not a BH but its metric is stationary, axisymmetric,
reflection symmetric and asymptotically flat, then the waves will likely be
triperiodic, as for a BH. We suggest that the evolutions of the waves' three
fundamental frequencies and of the complex amplitudes of their spectral
components encode (in principle) details of the central body's metric, the
energy and angular momentum exchange between the central body and the orbit,
and the time-evolving orbital elements. We estimate that advanced ground-based
detectors can constrain central body deviations from a BH with interesting
accuracy.
\end{abstract}

\maketitle

First-generation interferometric gravitational-wave (GW) detectors, such as
such as Laser Interferometer Gravitational-Wave Observatory
(LIGO)~\cite{Barish:1999} and Virgo~\cite{Acernese:2006}, are searching for
GWs at or near their design sensitivities. In the next decade, Advanced LIGO
(AdvLIGO)~\cite{Fritschel:2003qw} and its international partners will increase
the volume of the Universe searched a thousandfold or more. The most promising
GW sources for this network are the inspiral and coalescence of black hole
(BH) and/or neutron star (NS) binaries. Current inspiral searches target
sources with total mass $M \lesssim 40 M_\odot$: NS binaries with masses
$1$--$3\,M_\odot$, BH binaries with masses $3$--$40\,M_\odot$, and NS-BH
binaries with components in these mass
ranges~\cite{Abbott:2005pe,Abbott:2005kq}. 

Ultraluminous x-ray observations and simulations of globular cluster dynamics
suggest the existence of intermediate mass black holes (IMBHs) with masses $M
\sim 10^2$--$10^4\,M_\odot$~\cite{Miller:2003sc}. The GWs from the inspiral of
a NS or stellar-mass BH into an IMBH with mass $M \sim 50$--$350 M_\odot$ will
lie in the frequency band of AdvLIGO. These \emph{intermediate-mass-ratio
inspirals} (IMRIs) are analogous to the extreme-mass-ratio inspirals (EMRIs)
of $\sim 10 M_\odot$ objects spiraling into $\sim 10^6 M_\odot$ BHs, targeted
by the planned LISA observatory~\cite{Gair:2004iv}. We consider NSs and BHs,
as less compact objects (e.g. white dwarfs) are tidally disrupted at
frequencies too low to be detectable in AdvLIGO.  

If we consider the possibility that the central body of an IMRI (or EMRI) is
not a black hole, but some other general relativistic object (e.g.~a boson
star or a naked singularity~\cite{Ryan:1997}), then we can quantify the
accuracy with which it has the properties predicted for a black hole that: (i)
it obeys the black-hole no-hair theorem (its spacetime geometry is the Kerr
metric, fully determined by its mass and spin), and (ii) its {\it tidal
coupling} (tide-induced transfer of energy and angular momentum between orbit
and body) agrees with black-hole predictions.  Searching for non-BH objects
may yield an unexpected discovery. 

We report on our initial explorations of the prospects for detecting GWs from
IMRIs and probing the properties of the IMRIs' central bodies.  We report on:
(i) IMRI event rate estimates in AdvLIGO, (ii) estimates of the efficacy of GW
template families for IMRI searches, (iii) explorations of the character of
the IMRI (EMRI) waves if the central body is not a black hole, (iv)
generalizations of Ryan's theorem concerning the information about the central
body carried by IMRI and EMRI waves, and (v) estimates of the accuracies with
which information can be extracted by AdvLIGO from IMRIs.

{\bf Event Rates for IMRIs with an IMBH central body.} We (Mandel 
{\it et al.}~\cite{BGMM:2006}) estimate that for IMBH spins $\chi \equiv $~spin
angular momentum$/M^2 \lesssim 0.3$, the distance ({\it range}) $R$ in Mpc to
which a network of three $4$~km AdvLIGO detectors could see IMRIs at a network
signal-to-noise ratio (SNR) of 8 is  \begin{eqnarray} R &\approx& \left[1 +
(\chi^2/2) (M/100M_\odot)^{1.5}\right]\sqrt{m/M_\odot} \times \nonumber\\
&&\left[800 - 540(M/100M_\odot) + 107(M/100M_\odot)^2\right]\nonumber.
\end{eqnarray}   (For IMBHs grown by minor mergers, typical spins will be
$\chi \sim \sqrt{m/M} \sim 0.2$, with few if any above $\sim 0.4$.)

Core-collapsed globular clusters are the most likely locations for IMRIs; they
may contain an IMBH and a high density of stellar mass BHs and
NSs~\cite{Miller:2003sc}.  Simulations show that it is possible to grow IMBHs
with masses up to  $M_{\rm max} \sim 350\ M_\odot$ through a series of mergers
in the core of a cluster~\cite{O'Leary:2005tb}. Phinney~\cite{Phinney:2005}
estimates an upper limit on the IMRI rate in globular clusters as follows:
assume each cluster has an IMBH that grows from $\sim 50 M_\odot$ to $\sim 350
M_\odot$ by capturing objects of mass $m$ in $10^{10}$ years.  Core-collapsed
clusters have a space density of $0.7\ \mathrm{Mpc}^{-3}$, which gives an
estimated IMRI rate of $\sim 0.7 \times (300 M_\odot/m) \times 10^{-10}\
\mathrm{Mpc}^{-3} \mathrm{yr}^{-1}$. This leads to a limit of $\sim 10$ IMRI
detections per year in AdvLIGO.

A kick velocity $V_\mathrm{kick} > 50$~km/s will eject the merged black hole
from the cluster, placing an upper limit on $m$ of $m/M \lesssim 0.08$
($V_\mathrm{kick}$ depends on the symmetric mass ratio $\eta=mM/(m+M)^2$ as
$V_\mathrm{kick} \approx 12000 \eta^2 \sqrt{1-4\eta}
(1-0.93\eta)$~km/s~\cite{Gonzalez:2006}).  Black holes with masses $m \gtrsim
10 M_\odot$ will likely merge with the IMBH or be ejected from the core in
under $10^{10}$ years.  An estimate based on the dynamics of binary hardening
via 3-body interactions yields a rate of one detection per three years for
NS--IMBH inspirals or ten detections per year for BH--IMBH
inspirals~\cite{BGMM:2006}.  Optimizing AdvLIGO sensitivity at low frequencies
could improve these rates by a factor of $\sim 3$. For Initial
LIGO~\cite{Barish:1999}, rates are much lower due to lower detector
sensitivity and seismic noise below $40$~Hz, reducing $M_{\rm max}$ to
$\lesssim 100 M_\odot$. We estimate an IMRI rate in current detectors of $<
1/1000\ \textrm{yr}^{-1}$. 

{\bf Search Templates for IMRI Waves with an IMBH central body.} Matched
filter searches require templates of sufficient accuracy that the mismatch
between template and signal does not cause a large loss in event rate.  The
most accurate IMRI templates currently available come from BH perturbation
theory via numerical solution of the Teukolsky equation~\cite{Hughes:2001jr}.
Post-Newtonian (PN) templates~\cite{Kidder:1993,Blanchet:2002} and PN
approximations to Teukolsky waveforms~\cite{Tagoshi:1996gh} are inadequate
becuase IMRIs enter the detector frequency band when the binary separation is
$r \lesssim 15M$ and the PN expansion is poor.

Inspiral waveforms from black-hole perturbation theory are known only to first
order in $\eta$ plus ${\rm O}(\eta^2)$ in radiation reaction. It is important
to determine the effect of conservative finite-mass-ratio corrections ${\rm
O}(\eta^2)$, but tools to study these are not yet in hand.  We (Brown
\cite{Brown:2006}) estimate these effects by computing the mismatch (for
AdvLIGO) between restricted PN stationary-phase templates containing all known
$\eta$ terms, and the same templates linearized in $\eta$ plus ${\rm
O}(\eta^2)$ radiation reaction (cut off at the IMRI's innermost stable
circular orbit); this is the fractional SNR loss due to using templates
linearized in $\eta$.  Mismatches are computed at each PN order between $1.0$
and $3.5$ inclusive.  For a $1.4\,M_\odot$ NS--$100\,M_\odot$ IMBH IMRI, the
mismatch is $\lesssim 30\%$ for $\chi < 0.8$, and $\lesssim 15\%$ for $\chi <
0.3$.  For IMRIs with a larger IMBH mass, the mismatch decreases, as expected.
By allowing the linearized PN waveforms to have mass parameters different from
those of the nonlinear PN waveforms, and minimizing the mismatch over these
parameters, mismatch falls to less than $10\%$ in all except the most rapidly
spinning cases~\cite{Brown:2006}. Therefore, it is reasonable to expect that
the Teukolsky waveforms will lose no more than 10\% of the SNR due to
linearization in $\eta$ (hence no more than a 30\% loss of event rate). For
detection, it will be worthwhile, but not essential, to improve Teukolsky
waveforms by incorporating nonlinear corrections, but accurate parameter
measurement will require improvements.

{\bf IMRI and EMRI Orbits and Waves; Tri-periodic vs.~Ergodic.} Here we
entertain the possibility that the central body is not a black hole. We assume
its external spacetime geometry is stationary, axially and reflection
symmetric and asymptoticaly flat (SARSAF) with metric in the form $ds^2 = -
\alpha^2 dt^2 +\varpi^2(d\phi - \omega dt)^2 + g_{\theta\theta} d\theta^2 +
g_{rr} dr^2$ and all coefficients independent of the Killing time $t$ and
axial angle $\phi$. If the spacetime initially is not axisymmetric, rotation
will make it non-stationary; then presumably GW emission drives it to
stationarity and axisymmetry on astrophysically small time-scales. Almost all
stationary, axially symmetric, self-gravitating objects studied
observationally or theoretically are reflection symmetric.

A SARSAF solution to the vacuum Einstein equations is determined uniquely by
two families of scalar multipole moments: \emph{mass moments} $M_0 \equiv M$,
$M_2$ (mass quadrupole moment), $M_4, \ldots$ ; and \emph{current moments}
$S_1$ (spin angular momentum), $S_3$, $S_5, \ldots$~\cite{Hansen:1974}.  For
the Kerr metric (describing astrophysical black holes), the moments are fully
determined by the mass $M$ and dimensionless angular momentum $\chi \equiv
S_1/M^2$ via $M_\ell + i S_\ell = M^{l+1} (i\chi)^\ell$; this is the no-hair
theorem. LISA plans to measure as many moments as possible, via EMRI waves,
and determine the accuracy with which each moment satisfies this Kerr formula;
AdvLIGO can do the same for IMRIs.

For EMRIs and IMRIs, the orbiting object moves along an orbit that is nearly a
geodesic of the background metric; gravitational radiation reaction drives it
slowly from one geodesic to another.  If the central body is a Kerr BH, then:
(i) each geodesic has three isolating integrals of the motion: energy $E$,
axial angular momentum $L_z$, and Carter constant $Q$ (and a fourth,
``trivial'' integral, the length of the orbit's tangent vector); (ii) the
emitted gravitational waves are tri-periodic with  $h^{\mu\nu}=\Re
\sum_{Pkmn}h^{\mu\nu}_{Pkmn} e^{i(k\Omega_\theta+m\Omega_\phi+n\Omega_r)t}$
(for integer values of $k,m,n$)~\cite{Drasco:2004}. Here $P=+,\times$ is the
polarization, and the three \emph{principal frequencies} $\Omega_\theta$,
$\Omega_\phi$, $\Omega_r$, in a precise but subtle sense, are associated with
the orbital motion in the polar ($\theta$), azimuthal ($\phi$) and radial
($r$) directions.  The fundamental frequencies and complex amplitudes evolve
with time as the orbit evolves through a sequence of geodesics.

If the the Carter constant is lost in SARSAF spacetimes, motion may be ergodic
rather than tri-periodic, which would make detection of the gravitational
waves difficult.  Gu\'{e}ron and Letelier \cite{GL:2002} have used
Poincar\'{e} maps to search for ergodic geodesics in the static ($S_\ell=0$)
Erez-Rosen metric and we (Gair, {\it et al.}~\cite{Gair:2006}) have carried out
similar studies for a variant of the stationary ($S_\ell \ne 0$) Manko-Novikov
metric~\cite{MN:1992}.  Both of these metrics have arbitrary mass quadrupole
moment $M_2$, and higher order moments fixed by $M_2$, $S_1$ and $M$.  The
Poincar\'{e} maps in these spacetimes reveal that there are geodesics at very
small radii $r \sim $ few $M$ that appear ergodic, but non at large radii.
We~\cite{Gair:2006} found such geodesics only for oblate ($M_2 < 0$)
pertubations of spacetimes with sping, but in the Erez-Rosen
case~\cite{Gair:2006},  ergodicity appears only for prolate ($M_2 > 0$)
perturbations.  Radiation reaction drives the evolution of energy and angular
momentum in a way that makes it unlikely that the apparently ergodic geodesics
could be encountered in the course of an inspiral~\cite{Gair:2006}.  For the
apparently non-ergodic (integrable) geodesics, the spatial coordinates are
multi-periodic functions of Killing time $t$ to a numerical accuracy of
$10^{-7}$, and a general argument~\cite{Li:2006} based on the structure of the
gravitational propagator shows that their gravitational waves will have the
same kind of tri-periodic form as for Kerr BHs.

There are three possible explanations for the presence of large-radius orbits
that appear integrable and small-radius orbits that appear ergodic in the same
spacetime: (i) The orbits are actually integrable and actually ergodic,
respectively.  (ii) All the orbits are ergodic, but at large radii they
appear integrable to numerical accuracy because of the
Kolmogorov-Arnol'd-Moser theorem~\cite{Tabor}. (iii) All the orbits are
integrable, but at small radii they are made to appear chaotic by some
ill-understood numerical instability.  It is important to learn which is the
case, but for EMRI and IMRI wave observations, apparent integrability (or
ergodicity) has the same observational implications as actual integrability
(or ergodicity).

{\bf Information Carried by IMRI and EMRI Waves; Generalizing Ryan's Theorem.}
What information about the central body is encoded in the waveforms? We shall
assume the waveforms to be tri-periodic.  In principle, a large amount of
information can be encoded in the time evolution of the waves' three
fundamental frequencies $\Omega_\theta(t)$, $\Omega_\phi(t)$, $\Omega_r(t)$
and their complex amplitudes $h_{Pkmn}(t)$. It has been speculated that these
encode, fully and separably, the values of all the central body's multipole
moments $\{M_\ell, S_\ell\}$ and hence its metric~\cite{Ryan:1995wh}, the
rates at which the orbiting object's tidal pull deposits energy and angular
momentum into the central body, $\dot{E}_{\rm body}$ and $\dot{L}_{\rm body}$
(\emph{tidal coupling}) \cite{Fang:2005}, and the orbit's semi-latus rectum
$p(t)$, eccentricity $e(t)$ and inclination angle $\iota(t)$ (which carry the
same information as the isolating integrals)~\cite{Li:2006a}. That this might
be so is suggested by a special case that Ryan~\cite{Ryan:1995wh} has studied.
A trivial extension of Ryan's theorem~\cite{Li:2006a} leads to the following
algorithm for extracting information from the waves. Observe the time-evolving
modulation frequencies as functions of the time-evolving fundamental frequency
$f= \Omega_\phi/\pi$. From these, deduce the functions $\Omega_A(\Omega_\phi)$
and thence $\Omega_A(v)$ for $A=\theta,r$; expand in powers of $v\equiv (M
\Omega_\phi)^{1/3} \simeq ($orbital velocity); and read out the moments
(redundantly) from the two expansions.  Then, knowing the moments and thence
the metric, use the geodesic equation to deduce $p(t)$ from $\Omega_\phi(t)$
and use wave-generation theory to deduce $e(t)$ and $\iota(t)$ from particular
modulation amplitudes, $h_{Pkmn}(t)$.

We have generalized Ryan's theorem to strongly elliptical but nearly
equatorial orbits (Li~\cite{Li:2006}), to include tidal coupling (Li and
Lovelace~\cite{Li:2006a}), and are working on further generalizations.  For
{\it strongly elliptical but nearly equatorial orbits} the three fundamental
frequencies are independent of $\iota$ at first order.  We expand these
frequencies $\Omega_A(M_\ell,S_\ell,e,p)$ (with $A=\theta,\phi,r$) in powers
of $1/p$, with coefficients that depend on $e$ and the moments. Suppose we
observe a series of $2N+1$ values of $(\Omega_\theta,\Omega_\phi,\Omega_r)$
(for any integer $N$) during the course of an inspiral. This gives us $6N+3$
numbers, from which we can read off (via an algorithm based on our expansions
of the fundamental frequencies~\cite{Li:2006}): (i) the time evolution of
$e(t)$ and $p(t)$ ($2N+1$ values of each), (ii) the lowest $N+1$ mass moments,
and (iii) the lowest $N$ current moments.  By observing the evolving
amplitudes of the orbital-precession-induced modulations encoded in
$h_{Pkmn}$, we can recover the time evolution of $\iota$. Hence, in principle,
we have a full description of the spacetime.  In practice the methods of
extracting the information are likely to be quite different from these
algorithms. 

In the absence of tidal coupling Ryan demonstrated that, for a nearly
circular, nearly equatorial orbit, the central body's moments are encoded not
only in the waves' modulations, but also in the phase evolution of the waves'
dominant harmonic $f= \Omega_\phi/\pi$. We have extended this analysis to
deduce the power being deposited in the central body by tidal coupling,
$\dot{E}_{\rm body}$~\cite{Li:2006a}. We assume the moments and metric have
been deduced from the precessional modulations and then use deviations from
the Ryan-theorem phase evolution to deduce $\dot{E}_{\rm body}$.  Following
Ryan, we quantify the waves' phase evolution by $\Delta N(t) \equiv f^2
/\dot{f} = d(\mathrm{number\ of\ wave\ cycles})/d\ln f$.  From this definition
of $\Delta N$, we infer the rate of change of orbital energy: $\dot{E}_{\rm
orb} = (dE_{\rm orb}/d\Omega_\phi)(\Omega_\phi^2/\pi\Delta N)$. All
(time-evolving) quantities on the right side can be deduced from observation
plus the geodesic equation (for $dE_{\rm orb}/d\Omega_\phi$).  From the
deduced metric and the frequency $f(t)$ we can compute the power radiated to
infinity $\dot{E}_\infty$; and thence by energy conservation we can deduce the
power being deposited in the central body $\dot{E}_{\rm body} = - \dot{E}_{\rm
orb} - \dot{E}_\infty$~\cite{Li:2006a}.  We can also infer the angular
momentum transferred tidally to the central body, $\dot{L}_{\rm body}$, via
$\dot{L}_{\rm body} = \dot{E}_{\rm body}/\Omega_\phi$ (valid for nearly
circular, nearly equatorial orbits).

The above argument assumes that we can compute $\dot{E}_\infty$ without
knowing the boundary conditions of the inspiral-induced metric perturbation at
the central body, since we do not know the nature of the central body \emph{a
priori.}  For highly compact central bodies (those deep inside the perturbing
field's ``effective potential'') this is true to high but not complete
accuracy. The effect of boundary conditions at the central body on the
inspiral phase evolution is communicated outward to infinity mainly at low
frequencies (the orbital frequency and its low-order harmonics), and these
perturbations have great difficulty penetrating through the effective
potential. If the spacetime metric is Kerr, we have shown that the influence
of the inner boundary condition on the energy radiated to infinity is $\delta
\dot{E}_\infty \sim v^{10} \dot{E}_\infty$~\cite{Li:2006a}---five orders
smaller in the linear velocity $v$ than the tidal coupling $\dot{E}_{\rm body}
\sim v^5 \dot{E}_\infty$~\cite{Tagoshi:1997}.  Thus, to high accuracy we can
deduce $\dot{E}_\infty$ and thence $\dot{E}_{\rm body}$ from observations,
without knowing the body's precise nature.

{\bf Measurement Accuracies for AdvLIGO.} We have estimated how accurately
AdvLIGO, via IMRI waves, can constrain deviations of the
central body's quadrupole moment $M_2$ (Brown~\cite{Brown:2006}) and tidal
coupling $\dot E_{\rm body}$ (Fang~\cite{Hua:thesis}) from those of a Kerr
black hole.  Absent the true waveforms, we used PN waveforms as
signals and templates.  This introduces systematic error, but we believe
our results are indicative of the accuracies achievable.  Our
source is the circular inspiral of a neutron star into a $100M_\odot$ IMBH
(under the assumption that radiation reaction has circularized the
orbit~\cite{BGMM:2006}).  The orbit is inclined to the hole's equatorial
plane,  to produce a modulation crucial for breaking degeneracy
between the IMBH spin $\chi$ and $M_2$ and $\dot E_{\rm body}$.
 
To investigate $M_2$, we used templates accurate to 3.5PN order in phase
evolution~\cite{Blanchet:2002} and 1.5PN in spin-orbit
coupling~\cite{Kidder:1993}, added the effects of quadrupole-monopole
interaction~\cite{Poisson:1998} to both the phase and the precessional
modulation and numerically mapped the ambiguity function of these signals. For
a NS--IMBH IMRI ($M_2 = -\chi^2M^3$) with spin $\chi = 0.8$ and SNR $\sim 10$,
we found AdvLIGO measurement errors $\Delta \ln M \sim 0.006, \Delta \ln \chi
\sim 0.02$, and $\Delta \ln M_2 \sim 0.6$. If the IMBH spin is $\chi = 0.3$,
the error increases to $\Delta \ln M \sim 0.01, \Delta \ln \chi \sim 0.3$, and
$\Delta \ln M_2 \sim 2$. The accuracy of measurement depends strongly on
binary orientation; larger precessional modulation reduces the
errors~\cite{Brown:2006}.  

We model tidal coupling as $\dot{E}_{\rm body} \equiv \epsilon \, \dot{E}_{\rm
BH}$, where $\dot{E}_{\rm BH}$ is the energy flow into a Kerr black
hole~\cite{Tagoshi:1997} and seek to measure deviations parameterized by
$\epsilon$.  We constructed precessing waveforms~\cite{ACST:1994}, with
orbital inspiral phase given by the 3.5 PN approximation of the Teukolsky
waveforms~\cite{Tagoshi:1996gh}, and modulation linearized in inclination
angle~\cite{Sasaki:1995}.  We restricted inclination angles to $\iota <
\pi/4$, fixed the direction to the source and the central body's spin
orientation, and used the Fisher matrix to estimate parameter measurement
accuracies. For a black-hole central body with spin $\chi = 0.8$ and SNR$=
10$, we could measure $\epsilon$ to $\Delta\ln \epsilon \sim 1$ to 2,
increasing to $\Delta\ln \epsilon \sim 30$ at $\chi = 0.3$.

While these errors are larger than hoped, (i) the quadrupole moment $|M_2|$ of
a boson star with $\chi = 0.3$ is expected to be in the range $15$ to $100 \,
\chi^2M^3$~\cite{Ryan:1997}, so AdvLIGO could readily identify such a central
body, (ii) for small spins $\dot{E}_{\rm BH} \simeq - {1\over4}\chi v^5
\dot{E}_\infty$, and hence for $\chi = 0.3$, the accuracy of measuring tidal
coupling is $\Delta \dot{E}_{\rm body} \sim 30 \times 0.001 \,
\dot{E}_\infty$, i.e. $3\%$ of the power radiated to infinity, an
interesting accuracy for central bodies with anomalously large $\dot E_{\rm
body}$, and (iii)  observing an IMRI in each of the three AdvLIGO detectors
increases the accuracy of parameter estimation quoted by a factor of
$\sqrt{3}$; including additional detectors, e.g. Advanced VIRGO, could improve
this further.  In practice, parameter estimation will be pursued using Markov
Chain Monte Carlo techniques~\cite{Rover:2006ni,Brown:2006}.

Our results suggest that AdvLIGO will be able to verify with modest but
interesting accuracy that an IMRI's central body is a black hole, and perform
searches for non-Kerr central bodies.  AdvLIGO's accuracies for probing the
central body are far worse than LISA's (as expected, due to the thousand-fold
fewer wave cycles), but AdvLIGO is likely to be operational some years before
LISA. Its studies of central bodies will be a valuable precursor to LISA's
EMRI science, and might possibly yield a big surprise.  

\enlargethispage*{1000pt}

We are grateful to Y.~Chen, T.~Creighton, C.~Cutler, S.~Drasco, C.~Miller,
Y.~Pan and S.~Phinney for discussions.  This work was supported in part by NSF
grants PHY-0099568, PHY-0601459, NASA grant NNG04GK98G, a grant from the
Brinson Foundation, and NSF cooperative agreement PHY-0107417.

\begin{enumerate}
\small
\expandafter\ifx\csname natexlab\endcsname\relax\def\natexlab#1{#1}\fi
\expandafter\ifx\csname bibnamefont\endcsname\relax
  \def\bibnamefont#1{#1}\fi
\expandafter\ifx\csname bibfnamefont\endcsname\relax
  \def\bibfnamefont#1{#1}\fi
\expandafter\ifx\csname citenamefont\endcsname\relax
  \def\citenamefont#1{#1}\fi
\expandafter\ifx\csname url\endcsname\relax
  \def\url#1{\texttt{#1}}\fi
\expandafter\ifx\csname urlprefix\endcsname\relax\def\urlprefix{URL }\fi
\providecommand{\bibinfo}[2]{#2}
\providecommand{\eprint}[2][]{\url{#2}}

\setlength{\parskip}{-4pt}

\bibitem[{\citenamefont{Barish and Weiss}(1999)}]{Barish:1999}
\bibinfo{author}{\bibfnamefont{B.~C.}~\bibnamefont{Barish}} \bibnamefont{and}
\bibinfo{author}{\bibfnamefont{R.} \bibnamefont{Weiss}},
  \bibinfo{journal}{Phys. Today} \textbf{\bibinfo{volume}{10}},
  \bibinfo{pages}{44} (\bibinfo{year}{1999}).

\bibitem[{\citenamefont{Acernese et~al.}(2006)}]{Acernese:2006}
\bibinfo{author}{\bibfnamefont{F.}~\bibnamefont{Acernese}} \bibnamefont{et~al.},
  \bibinfo{journal}{Class. Quant. Grav.} \textbf{\bibinfo{volume}{23}},
  \bibinfo{pages}{S635} (\bibinfo{year}{2006}).

\bibitem[{\citenamefont{Fritschel}(2003)}]{Fritschel:2003qw}
\bibinfo{author}{\bibfnamefont{P.}~\bibnamefont{Fritschel}}
  (\bibinfo{year}{2003}), \eprint{gr-qc/0308090}.

\bibitem[{\citenamefont{Abbott et~al.}(2005)}]{Abbott:2005pe}
\bibinfo{author}{\bibfnamefont{B.}~\bibnamefont{Abbott}} \bibnamefont{et~al.},
  \bibinfo{journal}{Phys. Rev. D} \textbf{\bibinfo{volume}{72}},
  \bibinfo{pages}{082001} (\bibinfo{year}{2005}).

\bibitem[{\citenamefont{Abbott et~al.}(2006)}]{Abbott:2005kq}
\bibinfo{author}{\bibfnamefont{B.}~\bibnamefont{Abbott}} \bibnamefont{et~al.},
  \bibinfo{journal}{Phys. Rev. D} \textbf{\bibinfo{volume}{73}}
  \bibinfo{pages}{062001} (\bibinfo{year}{2006}).

\bibitem[{\citenamefont{Miller and Colbert}(2004)}]{Miller:2003sc}
\bibinfo{author}{\bibfnamefont{M.~C.} \bibnamefont{Miller}} \bibnamefont{and}
  \bibinfo{author}{\bibfnamefont{E.~J.~M.} \bibnamefont{Colbert}},
  \bibinfo{journal}{Int. J. Mod. Phys. D} \textbf{\bibinfo{volume}{13}},
  \bibinfo{pages}{1} (\bibinfo{year}{2004}).

\bibitem[{\citenamefont{Gair et~al.}(2004)}]{Gair:2004iv}
\bibinfo{author}{\bibfnamefont{J.~R.} \bibnamefont{Gair}} \bibnamefont{et~al.}
  \bibinfo{journal}{Class. Quant. Grav.} \textbf{\bibinfo{volume}{21}},
  \bibinfo{pages}{S1595} (\bibinfo{year}{2004}).

\bibitem[{\citenamefont{Ryan}(2007)}]{Ryan:1997}
\bibinfo{author}{\bibfnamefont{F.~D.}~\bibnamefont{Ryan}},
  \bibinfo{journal}{Phys. Rev. D} \textbf{\bibinfo{volume}{55}},
  \bibinfo{pages}{6081} (\bibinfo{year}{1997}).

\bibitem[{\citenamefont{O'Leary et~al.}(2005)}]{O'Leary:2005tb}
\bibinfo{author}{\bibfnamefont{R.~M.} \bibnamefont{O'Leary}}
  \bibnamefont{et~al.}, \bibinfo{journal}{Ap. J.}
  \textbf{\bibinfo{volume}{637}}, \bibinfo{pages}{937} (\bibinfo{year}{2005}).

\bibitem[{\citenamefont{Phinney}(2005)}]{Phinney:2005}
\bibinfo{author}{\bibfnamefont{E.~S.} \bibnamefont{Phinney}}
  (\bibinfo{year}{2005}), \bibinfo{note}{private communication}.

\bibitem[{\citenamefont{{Gonz\'{a}lez} et~al.}(2006)}]{Gonzalez:2006}
\bibinfo{author}{\bibfnamefont{J.~A.} \bibnamefont{{Gonz\'{a}lez}}}
  \bibnamefont{et~al.} (\bibinfo{year}{2006}), \eprint{gr-qc/0610154}.

\bibitem[{\citenamefont{{Brown} et~al.}(2006)\citenamefont{{Mandel}, {Brown}, 
  {Gair}, and {Miller}}}]{BGMM:2006}
\bibinfo{author}{\bibfnamefont{I.}~\bibnamefont{{Mandel}}},
  \bibinfo{author}{\bibfnamefont{D.~A.} \bibnamefont{{Brown}}},
  \bibinfo{author}{\bibfnamefont{J.~R.} \bibnamefont{{Gair}}}, \bibnamefont{and}
  \bibinfo{author}{\bibfnamefont{M.~C.} \bibnamefont{{Miller}}},
  \bibinfo{note}{Submitted to Ap.~J.}, \eprint{astro-ph/0705.0285}.

\bibitem[{\citenamefont{Hughes}(2001)}]{Hughes:2001jr}
\bibinfo{author}{\bibfnamefont{S.~A.} \bibnamefont{Hughes}},
  \bibinfo{journal}{Phys. Rev. D} \textbf{\bibinfo{volume}{64}},
  \bibinfo{pages}{064004} (\bibinfo{year}{2001}).

\bibitem[{\citenamefont{{Kidder} et~al.}(1993)}]{Kidder:1993}
\bibinfo{author}{\bibfnamefont{L.~E.} \bibnamefont{{Kidder}}}
  \bibnamefont{et~al.}, \bibinfo{journal}{Phys. Rev. D} \textbf{\bibinfo{volume}{47}},
  \bibinfo{pages}{R4183} (\bibinfo{year}{1993}).

\bibitem[{\citenamefont{{Blanchet} et~al.}(2002)}]{Blanchet:2002}
\bibinfo{author}{\bibfnamefont{L.}~\bibnamefont{{Blanchet}}}
  \bibnamefont{et~al.}, \bibinfo{journal}{Phys. Rev. D} \textbf{\bibinfo{volume}{65}},
  \bibinfo{pages}{061501(R)} (\bibinfo{year}{2002}).

\bibitem[{\citenamefont{Tagoshi et~al.}(1996)}]{Tagoshi:1996gh}
\bibinfo{author}{\bibfnamefont{H.}~\bibnamefont{Tagoshi}} \bibnamefont{et~al.},
  \bibinfo{journal}{Phys. Rev. D} \textbf{\bibinfo{volume}{54}},
  \bibinfo{pages}{1439} (\bibinfo{year}{1996}).

\bibitem[{\citenamefont{{Brown}}(2006)}]{Brown:2006}
\bibinfo{author}{\bibfnamefont{D.~A.} \bibnamefont{{Brown}}},
  \bibinfo{note}{in preparation}.

\bibitem[{\citenamefont{Hansen}(1974)}]{Hansen:1974}
\bibinfo{author}{\bibfnamefont{R.}~\bibnamefont{Hansen}}, \bibinfo{journal}{J.
  Math. Phys.} \textbf{\bibinfo{volume}{15}}, \bibinfo{pages}{24}
  (\bibinfo{year}{1974}).

\bibitem[{\citenamefont{Drasco and Hughes}(2004)}]{Drasco:2004}
\bibinfo{author}{\bibfnamefont{S.}~\bibnamefont{Drasco}} \bibnamefont{and}
  \bibinfo{author}{\bibfnamefont{S.~A.} \bibnamefont{Hughes}},
  \bibinfo{journal}{Phys. Rev. D} \textbf{\bibinfo{volume}{69}},
  \bibinfo{pages}{044015} (\bibinfo{year}{2004}).

\bibitem[{\citenamefont{Gu\'{e}ron and Letelier}(2002)}]{GL:2002}
\bibinfo{author}{\bibfnamefont{E.}~\bibnamefont{Gu\'{e}ron}} \bibnamefont{and}
  \bibinfo{author}{\bibfnamefont{P.~S.} \bibnamefont{Letelier}},
  \bibinfo{journal}{Phys. Rev. E} \textbf{\bibinfo{volume}{66}},
  \bibinfo{pages}{046611} (\bibinfo{year}{2002}).

\bibitem[{\citenamefont{{Gair} et~al.}(2006)\citenamefont{{Gair}, {Li},
  {Lovelace}, {Mandel}, and {Fang}}}]{Gair:2006}
\bibinfo{author}{\bibfnamefont{J.~R.} \bibnamefont{{Gair}}},
  \bibinfo{author}{\bibfnamefont{C.}~\bibnamefont{{Li}}},
  \bibinfo{author}{\bibfnamefont{G.}~\bibnamefont{{Lovelace}}},
  \bibinfo{author}{\bibfnamefont{I.}~\bibnamefont{{Mandel}}}, \bibnamefont{and}
  \bibinfo{author}{\bibfnamefont{H.}~\bibnamefont{{Fang}}},
  \bibinfo{note}{in preparation}.

\bibitem[{\citenamefont{Manko and Novikov}(1992)}]{MN:1992}
\bibinfo{author}{\bibfnamefont{V.~S.} \bibnamefont{Manko}} \bibnamefont{and}
  \bibinfo{author}{\bibfnamefont{I.~D.} \bibnamefont{Novikov}},
  \bibinfo{journal}{Class. Quant. Grav.} \textbf{\bibinfo{volume}{9}},
  \bibinfo{pages}{2477} (\bibinfo{year}{1992}).

\bibitem[{\citenamefont{{Li}}(2006)}]{Li:2006}
\bibinfo{author}{\bibfnamefont{C.}~\bibnamefont{{Li}}},
  \bibinfo{note}{in preparation}.

\bibitem[{\citenamefont{Tabor}(1989)}]{Tabor}
\bibinfo{author}{\bibfnamefont{M.}~\bibnamefont{Tabor}},
  \emph{\bibinfo{title}{Chaos and Integrability in nonlinear Dynamics}}
  (\bibinfo{publisher}{John Wiley \& Sons Inc.}, \bibinfo{year}{1989}).

\bibitem[{\citenamefont{Ryan}(1995)}]{Ryan:1995wh}
\bibinfo{author}{\bibfnamefont{F.~D.} \bibnamefont{Ryan}},
  \bibinfo{journal}{Phys. Rev. D} \textbf{\bibinfo{volume}{52}},
  \bibinfo{pages}{5707} (\bibinfo{year}{1995}).

\bibitem[{\citenamefont{Fang and Lovelace}(2005)}]{Fang:2005}
\bibinfo{author}{\bibfnamefont{H.}~\bibnamefont{Fang}} \bibnamefont{and}
  \bibinfo{author}{\bibfnamefont{G.}~\bibnamefont{Lovelace}},
  \bibinfo{journal}{Phys. Rev. D} \textbf{\bibinfo{volume}{72}},
  \bibinfo{pages}{124016} (\bibinfo{year}{2005}).

\bibitem[{\citenamefont{{Li} et~al.}(2006)\citenamefont{{Li}, and 
  {Lovelace}}}]{Li:2006a}
\bibinfo{author}{\bibfnamefont{C.}~\bibnamefont{{Li}}},
  \bibnamefont{and}
  \bibinfo{author}{\bibfnamefont{G.}~\bibnamefont{{Lovelace}}},
  \bibinfo{note}{Submitted to Phys. Rev. D},
  \eprint{gr-qc/0702146}.

\bibitem[{\citenamefont{{Tagoshi} et~al.}(1997)}]{Tagoshi:1997}
\bibinfo{author}{\bibfnamefont{H.}~\bibnamefont{{Tagoshi}}}
  \bibnamefont{et~al.}, \bibinfo{journal}{Prog. Theor. Phys.}
  \textbf{\bibinfo{volume}{98}}, \bibinfo{pages}{829} (\bibinfo{year}{1997}).

\bibitem[{\citenamefont{Fang}(2006)}]{Hua:thesis}
\bibinfo{author}{\bibfnamefont{H.}~\bibnamefont{Fang}}, Ph.D. thesis,
  \bibinfo{school}{Caltech}, \bibinfo{address}{Pasadena, CA}.

\bibitem[{\citenamefont{{Poisson}}(1998)}]{Poisson:1998}
\bibinfo{author}{\bibfnamefont{E.}~\bibnamefont{{Poisson}}},
  \bibinfo{journal}{Phys. Rev. D} \textbf{\bibinfo{volume}{57}}, \bibinfo{pages}{5287}
  (\bibinfo{year}{1998}).

\bibitem[{\citenamefont{{Apostolatos} et~al.}(1995)}]{ACST:1994}
\bibinfo{author}{\bibfnamefont{T. A. }~\bibnamefont{{Apostolatos}}}
  \bibnamefont{et~al.}, \bibinfo{journal}{Phys. Rev. D}
  \textbf{\bibinfo{volume}{49}}, \bibinfo{pages}{6274} (\bibinfo{year}{1994}).
 
\bibitem[{\citenamefont{{Sasaki} et~al.}(1995)}]{Sasaki:1995}
\bibinfo{author}{\bibfnamefont{M.}~\bibnamefont{{Shibata}}}
  \bibnamefont{et~al.}, \bibinfo{journal}{Phys. Rev. D}
  \textbf{\bibinfo{volume}{51}}, \bibinfo{pages}{1646} (\bibinfo{year}{1995}).

\bibitem[{\citenamefont{{Rover} et~al.}(2006)}]{Rover:2006ni}
\bibinfo{author}{\bibfnamefont{C.}~\bibnamefont{{Rover}}}
  \bibnamefont{et~al.}, \bibinfo{journal}{Class. Quant. Grav.}
  \textbf{\bibinfo{volume}{23}}, \bibinfo{pages}{4895} (\bibinfo{year}{2006}).

\end{enumerate}

\end{document}